\documentclass[12pt]{iopart}
\usepackage{graphicx,cite}
\begin{document}

\title[On the 1D hydrogen atom with minimal length uncertainty]
{A note on the one-dimensional hydrogen atom with minimal length uncertainty}

\author{Pouria Pedram}
\address{Department of Physics, Science and Research Branch, Islamic Azad University, Tehran, Iran} \ead{p.pedram@srbiau.ac.ir}

\begin{abstract}
We present exact energy spectrum and eigenfunctions of the
one-dimensional hydrogen atom in the presence of the minimal length
uncertainty. By requiring the self-adjointness property of the
Hamiltonian, we completely determine the quantization condition. We
indicate that the single-valuedness criteria of the eigenfunctions
in non-deformed case is an emergent condition and the semiclassical
solutions exactly coincide with the quantum mechanical results. The
behavior of the wave functions at the origin in coordinate space and
in quasiposition space is discussed finally.
\end{abstract}

 \vspace{2pc}

\section{Introduction}
Studying the effects of the generalized (gravitational) uncertainty
principle (GUP) on various physical systems has attracted much
attention in recent years and many papers have appeared in
literature to address the modification of the Hamiltonians and their
energy spectrum and eigenfunctions in the presence of the minimal
length uncertainty. Indeed, this idea arises naturally from various
candidates of quantum gravity such as string theory
\cite{1,2,3-1,3-2,3-3,4}, loop quantum gravity \cite{5},
noncommutative spacetime \cite{6,7,8}, and black-holes gedanken
experiments \cite{9,10}. All these studies imply a finite lower
bound to the possible resolution of length of the order of the
Planck length $\ell_{Pl}=\sqrt{\frac{G\hbar}{c^3}}\approx 10^{-35}m$
where $G$ is Newton's gravitational constant.

The problem of the hydrogen atom is studied in \cite{11,12,13,14,15}
and the exact energy eigenvalues and eigenfunctions are obtained. In
the presence of the minimal length, Akhoury and Yao \cite{Akhoury},
and Bouaziz and Ferkous \cite{Bouaziz} have solved this problem for
zero angular momentum states and found exact expressions for the
GUP-corrected solutions. This problem is also studied numerically
and perturbatively in Refs.~\cite{Brau,Benczik}. Fityo \etal
detected a single-valuedness problem in Ref.~\cite{Akhoury} and
tried to present the quantization condition in  one-dimension by
requiring the symmetricity of the inverse of the position operator
on the eigenfunctions \cite{Tkachuk}. However, since there is a free
parameter in their solution, the energy spectrum is not completely
determined.

Here, using an alternative representation of the deformed algebra,
we find the quantization condition by requiring the self-adjointness
property of the Hamiltonian. Also we show that the validity of the
single-valuedness criteria for one-dimensional hydrogen atom in
ordinary quantum mechanics is an emergent condition. The
quasiposition space solutions, coordinate space solutions,
semiclassical solutions, and the validity of WKB approximation are
discussed finally.

\section{The generalized uncertainty principle}
Consider the following one-dimensional deformed commutation relation
\begin{eqnarray}\label{gupc}
[X,P]=\rmi\hbar(1+\beta P^2),
\end{eqnarray}
where for $\beta=0$ we recover the well-known commutation relation
in ordinary quantum mechanics. To exactly satisfy the above algebra,
Kempf, Mangano and Mann (KMM) have proposed the following
representation \cite{7}:
\begin{eqnarray}\label{k1}
X &=& (1+\beta p^2)x,\\
P &=& p,\label{k2}
\end{eqnarray}
where $x$ an $p$ are canonical position and momentum operators,
i.e., $[x,p]=i\hbar$. In fact, $X$ and $P$ are symmetric operators
on the dense domain $S_{\infty}$ with respect to the following
scalar product:
\begin{eqnarray}
\langle\psi|\phi\rangle=\int_{-\infty}^{+\infty}\frac{\rmd
p}{1+\beta p^2}\psi^{*}(p)\phi(p).
\end{eqnarray}
However, this representation is not unique and using appropriate
canonical transformations, we can obtain an alternative
representation. For instance, consider the following exact
representation \cite{pedramPRD,pedramPLB}:
\begin{eqnarray}\label{x0p01}
X &=& x,\\ P &=&
\frac{\tan\left(\sqrt{\beta}p\right)}{\sqrt{\beta}}.\label{x0p02}
\end{eqnarray}
These operators are formally self-adjoint, i.e. $A=A^\dagger$ for
$A\in\{X,P\}$ and they are symmetric subject to the inner product
\begin{eqnarray}\label{inner}
\langle\psi|\phi\rangle=\int_{-\frac{\pi}{2\sqrt{\beta}}}^{+\frac{\pi}{2\sqrt{\beta}}}\mathrm{d}p\,\psi^{*}(p)\phi(p).
\end{eqnarray}
Moreover, the ordinary nature of the position operator is preserved
in this representation. In fact, both exact representations are
equivalent and they are related by the following canonical
transformation:
\begin{eqnarray}
X&\longrightarrow&\left[1+\arctan^2\left(\sqrt{\beta}P\right)\right]X,\\
P&\longrightarrow&\arctan\left(\sqrt{\beta}P\right)/\sqrt{\beta},
\end{eqnarray}
which transforms equations (\ref{x0p01}) and (\ref{x0p02}) to
equations (\ref{k1}) and (\ref{k2}) subjected to equation
(\ref{gupc}). In this representation, the completeness relation and
scalar product can be written as
\begin{eqnarray}\label{comp1}
\langle p'|p\rangle= \delta(p-p'),\\
\int_{-\frac{\pi}{2\sqrt{\beta}}}^{+\frac{\pi}{2\sqrt{\beta}}}\rmd
p\, |p\rangle\langle p|=1.\label{comp2}
\end{eqnarray}
Note that the operator $X$ is not a true self-adjoint operator.
Although its adjoint $X^{\dagger}=i\hbar\partial/\partial p$ has the
same formal expression, it acts on a different space of functions,
namely
\begin{eqnarray}\label{dx1}
{\cal D}(X)&=&\bigg\{\phi,X\phi\in{\cal
L}^2\left(\frac{-\pi}{2\sqrt{\beta}},\frac{+\pi}{2\sqrt{\beta}}\right);
\phi\left(\frac{+\pi}{2\sqrt{\beta}}\right)=\phi\left(\frac{-\pi}{2\sqrt{\beta}}\right)=0\bigg\},\\
{\cal D}(X^{\dagger})&=&\bigg\{\varphi,X^{\dagger}\varphi\in{\cal
L}^2\left(\frac{-\pi}{2\sqrt{\beta}},\frac{+\pi}{2\sqrt{\beta}}\right);\mbox{no
other restriction on }\varphi\bigg\}.\label{dx2}
\end{eqnarray}
To show this let us check the symmetricity of the position operator
as
\begin{eqnarray}
\left\langle X\varphi\bigg| \phi\right\rangle=\left\langle
\varphi\bigg| X\phi\right\rangle,
\end{eqnarray}
where $\phi$ and $\varphi$ belong to the domain of $X$ and
$X^\dagger$, respectively. This condition can be rewritten as
\begin{eqnarray}
-\rmi
\int_{-\frac{\pi}{2\sqrt{\beta}}}^{+\frac{\pi}{2\sqrt{\beta}}}&&\frac{\partial
\varphi^{*}(p)}{\partial p}\phi(p)\rmd p-\rmi
\int_{-\frac{\pi}{2\sqrt{\beta}}}^{+\frac{\pi}{2\sqrt{\beta}}}\varphi^*(p)\frac{\partial
\phi(p)}{\partial p}\rmd p
\nonumber\\
&&=-\rmi\varphi^*(p)\phi(p)\Big|_{p=\frac{+\pi}{2\sqrt{\beta}}}+\rmi\varphi^*(p)\phi(p)\Big|_{p=\frac{-\pi}{2\sqrt{\beta}}}=0,
\end{eqnarray}
which verifies (\ref{dx1}) and (\ref{dx2}).  As it is also shown in
\cite{Kempf2}, any operator $X$ which obeys the commutation relation
(\ref{gupc}) is not a true self-adjoint operator. Before proceed
further, let us categorize the operators in terms of their
self-adjointness properties as follows:
\begin{enumerate}
  \item Self-adjoint operators: $A=A^\dagger$ and
  $\mathcal{D}(A)=\mathcal{D}(A^\dagger)$.
  \item Symmetric operators: $\langle A\psi|\phi\rangle=\langle
  \psi|A\phi\rangle$ for all $\psi,\phi\in\mathcal{D}(A)$.
\end{enumerate}
Note that a self-adjoint operator is symmetric but its inverse is
not always true. However, the symmetricity of an operator is
sufficient to ensure the reality of the eigenvalues.

\section{Momentum representation}
Now consider the one-dimensional hydrogen atom eigenvalue problem:
\begin{eqnarray}\label{x1}
P^2\phi-\frac{\alpha}{X}\phi=E\phi,
\end{eqnarray}
where we set $\hbar=1=2m$ and we take $\displaystyle E=-\epsilon$.
In this representation we have $X\phi=\rmi\displaystyle
\frac{\partial \phi}{\partial p}$ and  the action of inverse
operator $1/X$ is expressed as
\begin{eqnarray}\label{X}
\frac{1}{X}\phi(p)=-\rmi\int_{-\frac{\pi}{2\sqrt{\beta}}}^{p}\phi(q)\rmd
q+c,\hspace{1cm}\frac{-\pi}{2\sqrt{\beta}}<p<\frac{+\pi}{2\sqrt{\beta}},
\end{eqnarray}
where, as we show below, $c$ should be a constant. Note that the
application of (\ref{X}) with $c=0$ leads to the existence of the
only trivial solution $\phi(p) = 0$ \cite{Tkachuk}. Also in the
absence of GUP the presence of $c$ corresponds to derivative
discontinuity of eigenfunctions at the origin in the coordinate
representation \cite{12}. By definition (\ref{X}) we obtain
\begin{eqnarray}\label{x11}
X \frac{1}{X}\phi&=&\phi,\\
  \frac{1}{X}X\phi&=&\phi+c,\label{x12}
\end{eqnarray}
and
\begin{eqnarray}\label{x13}
\left[X ,\frac{1}{X}\right]\phi=-c.
\end{eqnarray}

Similarly, since the adjoint of the position operator has the same
formal expression of the position operator, i.e.
$X^\dagger\phi=\rmi\displaystyle \frac{\partial \phi}{\partial p}$,
the action of the adjoint of $1/X$ is expressed as
\begin{eqnarray}\label{X2}
\left(\frac{1}{X}\right)^\dagger
\phi(p)=-\rmi\int_{-\frac{\pi}{2\sqrt{\beta}}}^{p}\phi(q)\rmd
q+c^*,\hspace{1cm}\frac{-\pi}{2\sqrt{\beta}}<p<\frac{+\pi}{2\sqrt{\beta}},
\end{eqnarray}
and therefore we find
\begin{eqnarray}
X^\dagger\left(\frac{1}{X}\right)^\dagger\phi=\phi,\label{x21}\\
\left(\frac{1}{X}\right)^\dagger  X^\dagger\phi=\phi+c^*,\label{x22}\\
\left[X^\dagger ,\left(\frac{1}{X}\right)^\dagger
\right]\phi=-c^*\label{x23}.
\end{eqnarray}

We now prove that $X^{-1}$ is not a linear operator. In a basis
which $X$ as a linear operator is diagonal, the formal operational
relation $X \frac{1}{X}=1$ (\ref{x11}) implies that if $X^{-1}$ is a
linear operator with a matrix representation, it is also diagonal.
So we have
\begin{eqnarray}
[X,X^{-1}]=0,
\end{eqnarray}
which apparently contradicts with equation (\ref{x13}). The same
argument also applies for $X^\dagger$. Explicitly we have
\begin{eqnarray}
\frac{1}{X}\left[\mu\phi(p)+\nu\varphi(p)\right]&=&-\rmi\mu\int_{-\frac{\pi}{2\sqrt{\beta}}}^{p}\phi(q)\rmd
q-\rmi\nu\int_{-\frac{\pi}{2\sqrt{\beta}}}^{p}\varphi(q)\rmd
q+c,\nonumber\\
&\ne&\mu\frac{1}{X}\phi(p)+\nu\frac{1}{X}\varphi(p),\label{x2x}
\end{eqnarray}
and a similar relation for $\left(\frac{1}{X}\right)^\dagger$. Thus
$c$ is a constant not a linear functional.

Equations (\ref{X}) and (\ref{X2}) now result in
\begin{eqnarray}
\left[\frac{1}{X}-\left(\frac{1}{X}\right)^\dagger\right]\phi=2\mathrm{Im}[c].
\end{eqnarray}
Notice that this relation is valid regardless of the actual nature
of $c$. Now since the momentum operator $P$ is Hermitian, i.e.
$P=P^\dagger$, the Hermicity nature of the Hamiltonian requires
$\frac{1}{X}=\left(\frac{1}{X}\right)^\dagger$ or
\begin{eqnarray}\label{imc}
\mathrm{Im}[c]=0.
\end{eqnarray}
As we shall see this condition completely determines the
quantization condition. Indeed, the presence of a free parameter in
the energy spectrum reported by Fityo \etal \cite{Tkachuk} is due to
the fact that they only implemented the symmetricity condition for
the Hamiltonian. Here, we impose an stronger condition i.e.
``self-adjointness" to fix the energy spectrum. To ensure the
self-adjointness of the Hamiltonian we will check the domains of the
Hamiltonian and its adjoint at the end of this section.

The Schr\"odinger equation in momentum space then reads
\begin{eqnarray}\label{Seq}
-\frac{\tan^2\left(\sqrt{\beta}p\right)}{\beta}\phi(p)-\rmi\alpha\int_{-\frac{\pi}{2\sqrt{\beta}}}^{p}\phi(q)\,\rmd
q+\alpha c=\epsilon\,\phi(p).
\end{eqnarray}
If we differentiate this equation with respect to $p$ we find
\begin{eqnarray}
\phi'(p)+\beta\frac{\frac{2\sec^2\left(\sqrt{\beta}p\right)\tan\left(\sqrt{\beta}p\right)}
{\sqrt{\beta}}+\rmi\alpha}{\tan^2\left(\sqrt{\beta}p\right)+\beta\epsilon}\phi(p)=0.
\end{eqnarray}
The solution is
\begin{eqnarray}\label{sol0}
\fl\phi(p)=\mathcal{A}\frac{2\beta\epsilon\cos^2\left(\sqrt{\beta}p\right)}{1+\beta\epsilon-\left(1-\beta\epsilon\right)\cos\left(2\sqrt{\beta}p\right)}
\exp\left[\frac{\rmi\alpha}{1-\beta\epsilon}\left(\beta
p-\frac{1}{\sqrt{\epsilon}}\arctan\left[\frac{\tan\left(\sqrt{\beta}p\right)}{\sqrt{\beta\epsilon}}\right]\right)\right],\hspace{.75cm}
\end{eqnarray}
where can be rewritten as
\begin{eqnarray}\label{sol}
\fl\phi(p)=\mathcal{A}\frac{2\beta\epsilon\cos^2\left(\sqrt{\beta}p\right)}{1+\beta\epsilon-\left(1-\beta\epsilon\right)\cos\left(2\sqrt{\beta}p\right)}
\exp\left[\frac{\rmi\alpha\beta
}{1-\beta\epsilon}p\right]\left[\frac{1-\rmi\frac{\tan\left(\sqrt{\beta}p\right)}{\sqrt{\beta\epsilon}}}
{1+\rmi\frac{\tan\left(\sqrt{\beta}p\right)}{\sqrt{\beta\epsilon}}}\right]^{\frac{\alpha}{2\sqrt{\epsilon}\left(1-\beta\epsilon\right)}},
\end{eqnarray}
where $\mathcal{A}$ is the normalization coefficient. Substituting
the expression for eigenfunctions (\ref{sol}) into the eigenvalue
equation (\ref{Seq}) results in
\begin{eqnarray}\label{c}
c=\frac{1}{\alpha}\,\,\lim_{p\rightarrow\frac{-\pi}{2\sqrt{\beta}}}\left(\frac{\tan^2\left(\sqrt{\beta}p\right)}{\beta}
+\epsilon\right)\phi(p)=\mathcal{A}\frac{\epsilon}{\alpha}\,\rme^{\frac{\rmi\pi\alpha}{2\sqrt{\epsilon}\left(1+\sqrt{\beta\epsilon}\right)}}.
\end{eqnarray}
So the probability density in momentum space reads
\begin{eqnarray}\label{pd}
|\phi(p)|^2=\left[\frac{2\mathcal{A}\beta\epsilon\cos^2\left(\sqrt{\beta}p\right)}
{1+\beta\epsilon-\left(1-\beta\epsilon\right)\cos\left(2\sqrt{\beta}p\right)}\right]^2.
\end{eqnarray}
Also the normalization coefficient is given by
\begin{eqnarray}
\mathcal{A}=\sqrt{\frac{2}{\pi}}\epsilon^{-1/4}\frac{
1+\sqrt{\beta\epsilon } }{\sqrt{\left( 1 + 2\sqrt{\beta\epsilon }
\right)} }.
\end{eqnarray}
The Hermicity condition (\ref{imc}) now implies
\begin{eqnarray}\label{sin}
\sin\left[\frac{\pi
\alpha}{2\sqrt{\epsilon}\left(1+\sqrt{\beta\epsilon}\right)}\right]=0,
\end{eqnarray}
which results in the following quantization condition:
\begin{eqnarray}\label{spec2}
\frac{\alpha}{2\left(\sqrt{\epsilon}+\sqrt{\beta}\epsilon\right)
}=n,\hspace{3cm}n=1,2,\ldots\,.
\end{eqnarray}
So the exact energy spectrum is given by
\begin{eqnarray}\label{enrg}
E_n=-\epsilon_n=-\frac{1}{4\beta}\left(1-\sqrt{1+\frac{2\alpha}{n}\sqrt{\beta}}\right)^2,\hspace{1cm}n=1,2,\ldots,
\end{eqnarray}
without any free parameter obtained in \cite{Tkachuk}. Using
equation (\ref{sin}), eigenfunctions also satisfy the following
condition:
\begin{eqnarray}\label{intP}
\int_{-\frac{\pi}{2\sqrt{\beta}}}^{+\frac{\pi}{2\sqrt{\beta}}}\phi(p)\rmd
p=\mathcal{A}\frac{2\epsilon}{\alpha}\sin\left[\frac{\pi
\alpha}{2\sqrt{\epsilon}\left(1+\sqrt{\beta\epsilon}\right)}\right]=0.
\end{eqnarray}

It is worth mentioning that to derive the energy spectrum
(\ref{enrg}) we did not use any explicit assumption about the actual
nature of $c$. The energy spectrum is based on the wave function
(\ref{sol}) and the Hermiticity condition (\ref{imc}) where both
relations are obtained regardless of the fact that $c$ is a linear
functional or a constant. Indeed our result can be used as a check
for the validity of each assumption. Since the unitary
transformation leads to the dependence of the energy spectrum on
arbitrary phase, it implies that $c$ is a constant in agreement with
the proof presented after equation (\ref{x23}).

To find the domains of the Hamiltonian and its adjoint, since the
operator $P^2=\tan^2\left(\sqrt{\beta}p\right)/\beta$ is obviously a
symmetric operator, we write the symmetricity condition for $1/X$ as
\begin{eqnarray}\label{weak}
\left\langle \frac{1}{X}\varphi\bigg| \phi\right\rangle=\left\langle
\varphi\bigg| \frac{1}{X}\phi\right\rangle,
\end{eqnarray}
where $\phi$ and $\varphi$ belong to the domain of $H$ and
$H^\dagger$, respectively. This condition using the explicit
expression for operator $1/X$ (\ref{X}) can be rewritten as
\begin{eqnarray}
\fl\rmi
\int_{-\frac{\pi}{2\sqrt{\beta}}}^{+\frac{\pi}{2\sqrt{\beta}}}\phi(p)&&\rmd
p\int_{-\frac{\pi}{2\sqrt{\beta}}}^{p}\varphi^*(q)\rmd
q+c^*\int_{-\frac{\pi}{2\sqrt{\beta}}}^{+\frac{\pi}{2\sqrt{\beta}}}\phi(p)\rmd
p\nonumber \\ &&=-\rmi
\int_{-\frac{\pi}{2\sqrt{\beta}}}^{+\frac{\pi}{2\sqrt{\beta}}}\varphi^*(p)\rmd
p\int_{-\frac{\pi}{2\sqrt{\beta}}}^{p}\phi(q)\rmd
q+c\int_{-\frac{\pi}{2\sqrt{\beta}}}^{+\frac{\pi}{2\sqrt{\beta}}}\varphi^*(p)\rmd
p.
\end{eqnarray}
Now the identity
\begin{eqnarray}
\fl\int_{-\frac{\pi}{2\sqrt{\beta}}}^{+\frac{\pi}{2\sqrt{\beta}}}f(p)\rmd
p\int_{-\frac{\pi}{2\sqrt{\beta}}}^{p}g(q)\rmd q=
\int_{-\frac{\pi}{2\sqrt{\beta}}}^{+\frac{\pi}{2\sqrt{\beta}}}g(p)\rmd
p\left[\int_{-\frac{\pi}{2\sqrt{\beta}}}^{+\frac{\pi}{2\sqrt{\beta}}}f(q)\rmd
q-\int_{-\frac{\pi}{2\sqrt{\beta}}}^{p}f(q)\rmd q\right],
\end{eqnarray}
implies
\begin{eqnarray}
\fl\rmi\int_{-\frac{\pi}{2\sqrt{\beta}}}^{+\frac{\pi}{2\sqrt{\beta}}}\phi(p)\rmd
p\int_{-\frac{\pi}{2\sqrt{\beta}}}^{+\frac{\pi}{2\sqrt{\beta}}}\varphi^*(q)\rmd
q+c^*\int_{-\frac{\pi}{2\sqrt{\beta}}}^{+\frac{\pi}{2\sqrt{\beta}}}\phi(p)\rmd
p-c\int_{-\frac{\pi}{2\sqrt{\beta}}}^{+\frac{\pi}{2\sqrt{\beta}}}\varphi^*(p)\rmd
p=0.
\end{eqnarray}
Therefore, using (\ref{intP}) we obtain
\begin{eqnarray}\label{con-sym}
\int_{-\frac{\pi}{2\sqrt{\beta}}}^{+\frac{\pi}{2\sqrt{\beta}}}\varphi^*(p)\rmd
p=0.
\end{eqnarray}
Now because of (\ref{intP}) and (\ref{con-sym}) the domains of $H$
and $H^\dagger$ coincide
\begin{eqnarray}
{\cal D}(H)={\cal D}(H^{\dagger})=\bigg\{\phi\in{\cal
D}_{\mathrm{max}}\left(\frac{-\pi}{2\sqrt{\beta}},\frac{+\pi}{2\sqrt{\beta}}\right);
\int_{-\frac{\pi}{2\sqrt{\beta}}}^{+\frac{\pi}{2\sqrt{\beta}}}\phi(p)\rmd
p=0\bigg\},
\end{eqnarray}
and the Hamiltonian is rendered a true self-adjoint operator
\cite{Bonneau}.

\subsection{Single-valuedness criteria}
The requirement of single-valuedness of the eigenfunctions
(\ref{sol}) leads to the following quantization condition:
\begin{eqnarray}\label{spec1}
\frac{\alpha}{2\sqrt{\epsilon}\left(1-\beta\epsilon\right)
}=m,\hspace{3cm}m=1,2,\ldots\,.
\end{eqnarray}
Note that unlike \cite{Tkachuk} there is no other term in
(\ref{sol}) that influences the single-valuedness criteria. But the
eigenfunctions obeying quantization condition (\ref{spec1}) do not
satisfy (\ref{intP}). Comparison between the two quantization
conditions (\ref{spec2}) and (\ref{spec1}) shows that
\begin{eqnarray}
m=\frac{n}{1-\sqrt{\beta\epsilon}}.
\end{eqnarray}
So the single-valuedness criteria of the eigenfunctions ($m\in$
Integers) is only valid at the limit $\beta\rightarrow0$, i.e., the
absence of GUP. In other words, in this case, the
``single-valuedness'' criteria can be considered as an emergent
condition rather than a fundamental one.

\subsection{Perturbative approach}
By expanding the energy spectrum in terms of the GUP parameter we
find
\begin{eqnarray}\label{per0}
E_n=-\frac{\alpha^2}{4n^2}+\frac{\alpha^3}{4n^3}\sqrt{\beta}-\frac{5\alpha^4}{16n^4}\beta+\mathcal{O}(\beta^{3/2}),\hspace{1cm}n=1,2,\ldots\,.
\end{eqnarray}
So the first correction to the energy spectrum is proportional to
$\sqrt{\beta}$. This result can be also understood from the
perturbative study of this problem. Consider the GUP-corrected
Hamiltonian to first-order of the deformation parameter
\begin{eqnarray}\label{per}
H\simeq p^2-\frac{\alpha}{x}+({2}/{3})\beta p^4.
\end{eqnarray}
The evaluation of the first-order energy spectrum leads to
\begin{eqnarray}
E_n=E_n^0+\Delta E_n,
\end{eqnarray}
where $E_n^0$ are unperturbed energy eigenvalues and $\Delta E_n$
are given by
\begin{eqnarray}\label{per2}
\Delta E_n=\frac{2}{3}\beta \left\langle
\phi^0_n(p)\big|p^4\big|\phi^0_n(p)\right\rangle.
\end{eqnarray}
Here $\phi_n^0(p)$ are solutions of (\ref{per}) with $\beta=0$. Also
equation (\ref{pd}) shows that at this limit the probability density
in momentum space takes the following form
\begin{eqnarray}
|\phi_n^0(p)|^2=\frac{2\epsilon^{3/2}}{\pi\left(p^2+\epsilon\right)^2}=\frac{4\alpha^3n}{\pi\left(\alpha^2+4n^2
p^2\right)^2}.
\end{eqnarray}
So the right hand side of (\ref{per2}) diverges that would explain
the $\sqrt{\beta}$ term in (\ref{per0}): the integral is linearly
divergent at large $p$, and since the natural momentum scale is
$1/\sqrt{\beta}$ the net result is of order $\beta \times
1/\sqrt{\beta}$ which gives $\sqrt{\beta}$ and this divergent behavior also comes from all
higher moments $\langle p^{2n}\rangle$ showing that expansion around
$\beta=0$ is not analytic.

\section{Quasiposition representation}
Following \cite{7} we define the maximal localization states
$|\phi^{\mathrm{ML}}_\xi\rangle$ with the following properties:
\begin{eqnarray}\label{miangin}
\langle \phi^{\mathrm{ML}}_\xi|X|\phi^{\mathrm{ML}}_\xi\rangle=\xi,
\end{eqnarray}
and
\begin{eqnarray}\label{deltaXML}
\Delta X_{|\phi^{\mathrm{ML}}_\xi\rangle}=(\Delta
X)_{min}=\hbar\sqrt{\beta}.
\end{eqnarray}
These states also satisfy
\begin{eqnarray}
\bigg(X-\langle X\rangle+\frac{\langle[X,P]\rangle}{2(\Delta
P)^2}\left(P-\langle P\rangle\right)\bigg)|\phi\rangle=0,
\end{eqnarray}
where $\langle[X,P]\rangle= \rmi\hbar\left(1+\beta (\Delta
P)^2+\beta\langle P\rangle^2 \right)$. So in momentum space the
above equation takes the form
\begin{eqnarray}
\fl\left[\rmi\hbar\frac{\partial}{\partial p}-\langle
X\rangle+\rmi\hbar\frac{1+\beta (\Delta P)^2+\beta\langle
P\rangle^2}{2(\Delta
P)^2}\left(\frac{\tan\left(\sqrt{\beta}p\right)}{\sqrt{\beta}}-\langle
P\rangle\right)\right]\phi(p)=0,
\end{eqnarray}
which has the solution
\begin{eqnarray}
\phi(p)= \mathcal{N}\exp\Bigg[&&\left(-\frac{\rmi}{\hbar}\langle
X\rangle+\frac{1+\beta (\Delta P)^2+\beta\langle
P\rangle^2}{2(\Delta P)^2}\langle P\rangle\right)p\nonumber\\
&&+\left(\frac{1+\beta (\Delta P)^2+\beta\langle
P\rangle^2}{2(\Delta P)^2}\right)
\frac{\ln\left[\cos\left(\sqrt{\beta}p\right)\right]}{\beta}\Bigg].
\end{eqnarray}
To find the absolutely maximal localization states we need to choose
the critical momentum uncertainty $\Delta P=1/\sqrt{\beta}$ that
gives the minimal length uncertainty and take $\langle P\rangle=0$,
i.e.,
\begin{eqnarray}\label{psiML}
\phi^{\mathrm{ML}}_\xi(p)=
\mathcal{N}\cos\left(\sqrt{\beta}p\right)\rme^{\frac{-\rmi
p\xi}{\hbar}},
\end{eqnarray}
where the normalization factor is given by
\begin{eqnarray}
\mathcal{N}=\sqrt{\frac{2\sqrt{\beta}}{\pi}}.
\end{eqnarray}
It is straightforward to check that $\phi^{\mathrm{ML}}_\xi(p)$
exactly satisfies (\ref{miangin}) and (\ref{deltaXML}). Because of
the fuzziness of space, these maximal localization states are not
mutually orthogonal.
\begin{eqnarray}
\fl\langle \phi^{\mathrm{ML}}_{\xi'}|\phi^{\mathrm{ML}}_\xi\rangle=
\mathcal{N}^2\int_{-\frac{\pi}{2\sqrt{\beta}}}^{+\frac{\pi}{2\sqrt{\beta}}}\rmd
p\cos^2\left(\sqrt{\beta}p\right)\rme^{\frac{-\rmi
p(\xi-\xi')}{\hbar}}=\frac{8\beta^{3/2}\hbar^3}{\pi}\frac{\sin\left[\frac{\pi(\xi-\xi')}{2\hbar\sqrt{\beta}}\right]}{(\xi-\xi')^3-4\beta\hbar^2(\xi-\xi')}.
\end{eqnarray}
To find the quasiposition wave function $\psi(\xi)$, we  define
\begin{eqnarray}
\psi(\xi)\equiv\langle\phi^{\mathrm{ML}}_\xi|\phi\rangle,
\end{eqnarray}
where in the limit $\beta\rightarrow0$ it goes to the ordinary
position wave function $\psi(\xi)=\langle\xi|\phi\rangle$. Now the
transformation of the wave function in the momentum representation
into its counterpart quasiposition wave function is
\begin{eqnarray}
\psi(\xi)=\mathcal{N}\int_{-\frac{\pi}{2\sqrt{\beta}}}^{+\frac{\pi}{2\sqrt{\beta}}}\rmd
p\cos\left(\sqrt{\beta}p\right)\rme^{\frac{\rmi
p\xi}{\hbar}}\phi(p).\label{psiF}
\end{eqnarray}
Although, regardless of energy, all wave functions in position space
vanish at the origin for $\beta=0$, i.e.~$\psi^0(0)=0$, in the
presence of the minimal length, the quasiposition wave functions do
not vanish generally at the origin where can be checked by numerical
evaluation of (\ref{psiF}).

\section{Coordinate representation}
The eigenfunctions of the position operator in momentum space are
given by the solutions of the eigenvalue equation
\begin{eqnarray}
X\,u_x(p)=x\,u_x(p),
\end{eqnarray}
where $u_x(p)=\langle p|x\rangle$. The normalized solutions are
\begin{eqnarray}
u_x(p)=\sqrt{\frac{\sqrt{\beta}}{\pi}}\exp\left({-\rmi\frac{
p}{\hbar}} x\right),
\end{eqnarray}
Now using (\ref{comp2}) we find the wave function in coordinate
space as
\begin{eqnarray}\label{foo}
\eta(x)=\sqrt{\frac{\sqrt{\beta}}{\pi}}
\int_{-\frac{\pi}{2\sqrt{\beta}}}^{+\frac{\pi}{2\sqrt{\beta}}}
\rme^{\frac{\rmi px}{\hbar}}\phi(p)\rmd p.
\end{eqnarray}
Note that since the eigenfunctions of the position operator satisfy
the zero uncertainty relation, i.e. $\Delta X_{|x\rangle}=0$,
$\eta(x)$ is not the physical wave function of the system. However,
the generalized Schr\"odinger equation in coordinate space has a
simple structure and $\eta(x)$ can be also considered as an
intermediate solution.

In coordinate space, the wave function at the origin is given by
\begin{eqnarray}
\eta(0)=\sqrt{\frac{\sqrt{\beta}}{\pi}}
\int_{-\frac{\pi}{2\sqrt{\beta}}}^{+\frac{\pi}{2\sqrt{\beta}}}
\phi(p)\rmd p.
\end{eqnarray}
Now because of (\ref{intP}) it is rendered zero, namely
\begin{eqnarray}
\eta(x)\Big|_{x=0}=0.
\end{eqnarray}
So the coordinate space wave functions satisfy the Dirichlet
boundary condition as well as in ordinary quantum mechanics
\cite{12}.

\section{Semiclassical approach}
The energy spectrum can be also
found using the Bohr-Sommerfeld quantization condition
\begin{eqnarray}
\oint p\,\rmd x=2n\pi,\hspace{3cm}n=1,2,\ldots\,.
\end{eqnarray}
The corresponding classical Hamiltonian to this system is
\begin{eqnarray}\label{eqP0}
H(x,p)=\frac{\tan^2\left(\sqrt{\beta}p\right)}{\beta}-\frac{\alpha}{x}.
\end{eqnarray}
Since the Hamiltonian is conserved, i.e. $H(x,p)=E$, we can express
$x$ as a function of $p$
\begin{eqnarray}
x=\frac{\alpha\beta}{\tan^2\left(\sqrt{\beta}p\right)-\beta E},
\end{eqnarray}
and use the identity $\oint p\,\rmd x=-\oint x\rmd p$. When the
particle leaves the origin in positive direction $p$ changes from
$+\frac{\pi}{2\sqrt{\beta}}$ to $0$ and when it returns to the
origin in negative direction $p$ changes from $0$ to
$-\frac{\pi}{2\sqrt{\beta}}$. So $\displaystyle-\oint x\rmd
p=\int_{-\frac{\pi}{2\sqrt{\beta}}}^{+\frac{\pi}{2\sqrt{\beta}}}x\rmd
p$ and for the negative energy bound states we find
\begin{eqnarray}
2n\pi=\int_{-\frac{\pi}{2\sqrt{\beta}}}^{+\frac{\pi}{2\sqrt{\beta}}}\frac{\alpha\beta}{\tan^2\left(\sqrt{\beta}p\right)-\beta
E}\rmd p=\frac{\pi\alpha}{\sqrt{\epsilon}+\sqrt{\beta}\epsilon},
\end{eqnarray}
which exactly agrees with the quantum mechanical result
(\ref{spec2}).

\section{WKB approximation} To check the validity of the
Bohr-Sommerfeld quantization rule for this modified quantum
mechanics, let us write the first-order generalized Schr\"odinger
equation corresponding to the Hamiltonian
\begin{eqnarray}\label{eqP}
H(x,p)=p^2+\frac{2}{3}\beta p^4+V(x).
\end{eqnarray}
as
\begin{eqnarray}
-\hbar^2\frac{\partial^2\psi(x)}{\partial x^2}+
\frac{2}{3}\hbar^4\beta\frac{\partial^4\psi(x)}{\partial
x^4}+V(x)\psi(x)=E\,\psi(x),
\end{eqnarray}
and take
\begin{eqnarray}
\psi(x)=\rme^{\rmi\Phi(x)},
\end{eqnarray}
where $\Phi(x)$ can be expanded as a power series in $\hbar$ in the
semiclassical approximation i.e.
\begin{eqnarray}
\Phi(x)=\frac{1}{\hbar}\sum_{n=0}^{\infty}\hbar^n\Phi_n(x).
\end{eqnarray}
So we have
\begin{eqnarray}
\frac{\partial^2\psi(x)}{\partial
x^2}&=&-\left(\Phi'^2-\rmi\Phi''\right)\psi(x),\\
\frac{\partial^4\psi(x)}{\partial x^4}&=&\left(\Phi'^4
-6\rmi\Phi'^2\Phi''-3\Phi''^2-4\Phi'''\Phi' +\rmi
\Phi''''\right)\psi(x),
\end{eqnarray}
where $\Phi'$ indicates the derivative of $\Phi$ with respect to
$x$. To the zeroth-order ($\Phi(x)\simeq\Phi_0(x)/\hbar$) and for
$\hbar\rightarrow0$ we obtain
\begin{eqnarray}
\Phi_0'^2+\frac{2\beta}{3}\Phi_0'^4=E-V(x).
\end{eqnarray}
Now the comparison with Eq.~(\ref{eqP}) shows $\Phi_0'=p$ and
consequently
\begin{eqnarray}
\psi(x)\simeq \exp\left[{\frac{\rmi}{\hbar}\int
p\,\mathrm{d}x}\right],
\end{eqnarray}
which is the usual zeroth-order WKB wave function obeying the
Bohr-Sommerfeld quantization rule. The generalization of this result
to higher order perturbed Hamiltonian and the non-perturbative
Hamiltonian (\ref{eqP0}) is also straightforward. Indeed, the
agreement between the exact and semiclassical results is the
manifestation of the validity of the Bohr-Sommerfeld quantization
rule in this modified quantum mechanics \cite{Fityo}.

\section{Conclusions}
In this paper, we found exact energy eigenvalues and eigenfunctions
of the one-dimensional hydrogen atom by requiring the
self-adjointness of the Hamiltonian. The energy spectrum is based on
two main relations. The first is the differential equation in
momentum space (\ref{sol}) and the second is the Hermicity condition
(\ref{imc}). In Ref.~\cite{Tkachuk} the differential equation is
solved in momentum space using KMM representation. However, the
Hermicity condition was not taken into account. Although the formal
expression of the solutions are different, the value of $c$ is
rendered to be similar to (\ref{c}), namely \cite{Tkachuk}
\begin{eqnarray}
c=\sqrt{\frac{2}{\pi}}\frac{\epsilon^{\frac{3}{4}}}{\alpha}
\frac{1+\sqrt{\beta\epsilon}}{\sqrt{1+2\sqrt{\beta\epsilon}}}
\exp\left[\frac{\rmi\alpha\pi}{2\left(\sqrt{\epsilon}+\sqrt{\beta}\epsilon\right)}\right].
\end{eqnarray}
So if instead of the weaker condition (\ref{weak}), we apply the
Hermicity condition (\ref{imc})  to the results of
Ref.~\cite{Tkachuk}, we recover the correct energy spectrum without
any free parameter. As stated before, the Hermicity condition is
hold whether we take $c$ to be constant or linear functional.
However, the algebraic structure of $X^{-1}$ [see e.g.
(\ref{x11})-(\ref{x13})] and the behavior of the solutions under the
unitary transformation do not support the latter assumption.

After finding the maximal localization states we obtained the
quasiposition wave functions and showed that unlike the coordinate
space solutions, they do not vanish generally at the origin.
Moreover, we indicated that the WKB approximation is valid in this
deformed algebra and the semiclassical energy spectrum exactly
coincides with the quantum mechanical results. It is also shown that
the single-valuedness criteria is an emergent condition in ordinary
quantum mechanics.

\ack I am very grateful to Rajesh Parwani, Taras Fityo, and Kourosh
Nozari for fruitful discussions and suggestions and for a critical
reading of the manuscript.

\end{document}